\ifdefined \preprintFormat
	\documentclass[aps,preprint,showpacs,superscriptaddress,groupedaddress]{revtex4-1} 
\else
	\documentclass[aps,prl,twocolumn,showpacs,superscriptaddress,groupedaddress]{revtex4-1} 
\fi
\usepackage{lineno}
\usepackage{graphicx} 
\usepackage{dcolumn} 
\usepackage{bm}        
\usepackage{amssymb} 
\usepackage{amsmath} 
\usepackage{stmaryrd}
\usepackage{fancyhdr}
\DeclareGraphicsExtensions{.pdf, .jpg, .png}

\begin{document}
    \title{Momentum chromospin classification of the fundamental fermions and bosons}
    \author{T. B. Edo}
    \email{tega.edo@sheffield.ac.uk}
    \affiliation{Department of Physics and Astronomy, University of Sheffield, S3 7RH, UK}
    \begin{abstract}
        We present a momentum chromospin scheme, the $\mathsf{KSC}$-{\it framework}, for parametrizing the elementary particles using the  underlying connection between their internal symmetries and their electromagnetic field boundaries. The linear momentum (or $\mathsf{K}$) chromospin branch of the framework deals with electric charge information, including the emergence of quantum chromodynamics color from fractional electric charges and its effect on quark confinement, while the angular momentum (or $\mathsf{S}$) chromospin branch of the framework develops the theory of spin to address the issue of three fermion generations.
    \end{abstract}
    \maketitle

	\section{I. Introduction}
    A major unresolved puzzle of the {\it Standard Model} (SM) is the duplication of matter particles into three generations. Earlier attempts at a resolution have employed  a wide range of ideas including complex octonions \cite{3G0}, catamorphy \cite{3G9}, extra-dimensional models \cite{3G1, 3G2, 3G3, 3G4, 3G5}, anomaly-free supersymmetric models \cite{3G6, 3G7, 3G8} and anthropic explanation \cite{3G10}. Here, we resolve this puzzle by parametrizing the various topological spin configurations of the elementary particles using the $\mathsf{KSC}$-{\it framework}. At the core of this framework is a space-like vector potential method---mutually exclusive to the time-like vector potential method of the SM \cite{qft1, qft2, qft3, qft4, qft5, qft6, SM1, SM2, SM3, SM4}---for extracting both the electromagnetic (EM) field and its source (charge or spin) from a particle's scalar potential. We also discuss the origin of quantum chromodynamics (QCD) color and its confinement property using its relation to the fractional electric charge of quarks and the structure of the quark's electric field. Although, several models have been proposed \cite{color2, color3, color4} that merge QCD color charge and the electric charge, they do not explain color confinement. Moreover, by combining the linear and angular momentum branches of the $\mathsf{KSC}$-{\it framework}, we discover the quantum mechanical process responsible for the phenomenon of flavor mixing in the lepton and quark sectors.

    \section{II. EM source boundary}
	The connection between the charge (spin) of the elementary particles and their linear (angular) momentum chromospin boundary parameter is best illustrated by the congruence of the EM field tensors and the group multiplication table of the momentum bases. The vectorial representation for the linear and angular momentum bases, which are consistent with the sign conventions of Maxwell's equations, are
	\begin{linenomath}
	    \begin{subequations}\label{eq:basesKS}
	        \begin{alignat}{2}
	            &{[\mathrm{K}_\mu]}^\gamma && \triangleq \mathrm{diag}(\begin{array}{cccc}  -1, & \,i, & \,i, & \,i \end{array}),\label{eq:basesK}\\
	            &{[\,\mathrm{S}_\mu\,]}^\gamma && \triangleq \mathrm{diag}(\begin{array}{cccc} ~~\,i, & 1, & 1, & 1 \end{array}),\label{eq:basesS}
	        \end{alignat}
	    \end{subequations}
	\end{linenomath}
    respectively. The dual bases, $\mathrm{K}^\mu$ and $\mathrm{S}^\mu$, are given by complex conjugation and Eq.~(\ref{eq:basesK}) imposes the following sign convention: $\partial^\mu=(-\partial_t,\partial_x,\partial_y,\partial_z)$ and $\partial_\mu=-(\partial_t,\partial_x,\partial_y,\partial_z)$.    
    \begin{table}[h]
        \caption{\label{tab:CayleyTable} Cayley table for the linear and angular momentum bases. $\mathrm{K}_t$ is the non-abelian group identity element and $\mathrm{S}_t$ swaps the momentum tag of the group element, i.e. $\mathrm{K} \rightleftharpoons \mathrm{S}$. This table provides a natural extension of Maxwell's equations to spin dynamics.}
		\begin{displaymath} 
			\begin{array}{ccc} 
			\otimes & {\mathrm{K}_\nu} & \mathrm{S}_\nu 
			\\\\ 
			{\mathrm{K}_\mu} 
			&\small{\begin{bmatrix} 
				-\mathrm{K}_t & -\mathrm{K}_x & -\mathrm{K}_y & -\mathrm{K}_z \\ 
				+\mathrm{K}_x & +\mathrm{K}_t & -\mathrm{S}_z & +\mathrm{S}_y \\ 
				+\mathrm{K}_y & +\mathrm{S}_z & +\mathrm{K}_t & -\mathrm{S}_x \\ 
				+\mathrm{K}_z & -\mathrm{S}_y & +\mathrm{S}_x & +\mathrm{K}_t \\ 
				\end{bmatrix}} 
			&\small{\begin{bmatrix} 
				-\mathrm{S}_t & -\mathrm{S}_x & -\mathrm{S}_y & -\mathrm{S}_z \\ 
				+\mathrm{S}_x & +\mathrm{S}_t & +\mathrm{K}_z & -\mathrm{K}_y \\ 
				+\mathrm{S}_y & -\mathrm{K}_z & +\mathrm{S}_t & +\mathrm{K}_x \\ 
				+\mathrm{S}_z & +\mathrm{K}_y & -\mathrm{K}_x & +\mathrm{S}_t \\ 
				\end{bmatrix}} 
			\\\\ 
			{\mathrm{S}_\mu} 
			&\small{\begin{bmatrix} 
				-\mathrm{S}_t & -\mathrm{S}_x & -\mathrm{S}_y & -\mathrm{S}_z \\ 
				+\mathrm{S}_x & +\mathrm{S}_t & +\mathrm{K}_z & -\mathrm{K}_y \\ 
				+\mathrm{S}_y & -\mathrm{K}_z & +\mathrm{S}_t & +\mathrm{K}_x \\ 
				+\mathrm{S}_z & +\mathrm{K}_y & -\mathrm{K}_x & +\mathrm{S}_t \\ 
				\end{bmatrix}} 
			&\small{\begin{bmatrix} 
				+\mathrm{K}_t & +\mathrm{K}_x & +\mathrm{K}_y & +\mathrm{K}_z \\ 
				-\mathrm{K}_x & -\mathrm{K}_t & +\mathrm{S}_z & -\mathrm{S}_y \\ 
				-\mathrm{K}_y & -\mathrm{S}_z & -\mathrm{K}_t & +\mathrm{S}_x \\ 
				-\mathrm{K}_z & +\mathrm{S}_y & -\mathrm{S}_x & -\mathrm{K}_t \\ 
				\end{bmatrix}} 
			\end{array} 
		\end{displaymath} 
    \end{table}
	We employ natural units where $\hbar=1$ and $c=1$. Table~\ref{tab:CayleyTable} shows the group multiplication table (or Cayley table) for the linear and angular momentum bases, and we employ a $4 \times 4$ matrix grouping to emphasize its role as the field bases for the standard EM tensor (top and bottom left matrices) and the Spin tensor (top and bottom right matrices). We thus infer that spin, like electric charge, has a concrete classical description. In algebraic form, the EM field bases for the charge and spin tensors are:
	\begin{linenomath}
	    \begin{subequations}\label{eq:tensorBases}
	        \begin{alignat}{6}
	            &\mathbb{F}_{\mu\nu}  &&=  \mathrm{K}_\mu &&\otimes \mathrm{K}_\nu &&= \mathrm{K}_\mu &&\times \mathrm{K}_\nu &&+\mathrm{K}_\mu \cdot \mathrm{K}_\nu,\label{eq:basesTensorF}\\
	            &\mathbb{M}_{\mu\nu} &&=  \mathrm{K}_\mu &&\otimes \mathrm{S}_\nu &&= \mathrm{K}_\mu &&\times \mathrm{S}_\nu &&+\mathrm{K}_\mu \cdot \mathrm{S}_\nu,\label{eq:basesTensorM}
	        \end{alignat}
	    \end{subequations}
	\end{linenomath}
    where we infer the form of the dot product---responsible for the source terms---from the Cayley table as: $\mathrm{K}_\mu \cdot \mathrm{K}_\nu = -\eta_{\mu\nu} \mathrm{K}_t $ and $\mathrm{K}_\mu \cdot \mathrm{S}_\nu = -\eta_{\mu\nu} \mathrm{S}_t $. The corresponding dual tensor bases are $\widetilde{\mathbb{F}}_{\mu\nu} = \mathrm{S}_\mu \times \mathrm{K}_\nu$ and $\widetilde{\mathbb{M}}_{\mu\nu} = \mathrm{S}_\mu \times \mathrm{S}_\nu$, where the absence of the dot product term implies that a real-valued vector potential cannot simultaneously encode charge and spin information. To get the field tensors, we substitute the derivative operator and the appropriate vector potential into Eq.~(\ref{eq:tensorBases}), i.e.
	\begin{linenomath}    
	    \begin{subequations}\label{eq:TensorFM}
	        \begin{alignat}{2}
	            &F^{\mu\nu}  &&=\partial^\mu A^\nu -\partial^\nu A^\mu +\eta^{\mu\nu}\partial_\gamma A^\gamma ,\label{eq:TensorF} \\
	            &M^{\mu\nu} &&= \partial^\mu{\tilde{A}}^\nu - \partial^\nu{\tilde{A}}^\mu + \eta^{\mu\nu}\partial_\gamma{\tilde{A}}^\gamma,\label{eq:TensorM}
	        \end{alignat}
	    \end{subequations}
	\end{linenomath}
    where $A^\mu$ and ${\tilde{A}}^\mu$ are the linear and angular momentum vector potentials respectively. The dual tensors are given by the usual relation: ${\widetilde{Y}}^{\alpha\beta} = \frac{1}{2}\varepsilon^{\alpha\beta\mu\nu}Y_{\mu\nu}$. The sign convention for the elements of the field tensors mirror their underlying field bases in Table~\ref{tab:CayleyTable}. Using the field tensors of Eq.~(\ref{eq:TensorFM}), we convert the inhomogeneous Maxwell equations into homogeneous wave equations for the vector potentials, namely
	\begin{linenomath}     
	    \begin{subequations}\label{eq:MaxwellEqns}
	        \begin{alignat}{3}
	            &\partial_\mu F^{\mu\nu}  &&= \partial^2A^\nu &&=0,\label{eq:MaxwellEqK} \\
	            &\partial_\mu M^{\mu\nu} &&= \partial^2{\tilde{A}}^\nu &&=0.\label{eq:MaxwellEqS}
	        \end{alignat}
	    \end{subequations}
	\end{linenomath}
    We derive Eq.~(\ref{eq:MaxwellEqns}) from the Lagrangian, $\mathcal{L}=E^{\mu}E_{\mu} - B^{\mu}B_{\mu}$, where the electric and magnetic fields are $E^t= \partial_\gamma A^\gamma$, $E^i = 2\partial^{[0} A^{i]} \oplus \varepsilon^{ijk}\partial^j{\tilde{A}}^k$ and $B^t= \partial_\gamma {\tilde{A}}^\gamma$, $B^i = 2\partial^{[0} \tilde{A}^{i]} \oplus \varepsilon^{ijk}\partial^j A^k$ respectively, and the incoherent sum $c = a \oplus b$ implies $c^2=a^2+b^2$. The charge and spin currents are
    \begin{linenomath} 
	    \begin{subequations}\label{eq:SourceKS}
	        \begin{align}
	            j_q^\mu &= -\partial^\mu E_t,\label{eq:SourceK} \\
	            j_s^\mu &= -\partial^\mu B_t,\label{eq:SourceS}
	        \end{align}
	    \end{subequations}
	\end{linenomath}
    respectively, with the continuity equation tensor: $\partial^\mu \otimes j^\nu (= 0)$. The singular form of a free particle's charge suggests that the revised Lorentz gauge condition for the linear momentum vector potential is $E_t=t\cdot\delta^{(\rho)}(\bm{r})$, where $\rho \in \{0\,..\,3\}$ encodes the magnitude of the particle's electric charge: $Q=(\rho/3)e$. Similarly, the revised Lorentz gauge condition for the angular momentum vector potential is $B_t=t\cdot{\delta'}^{(\sigma)}(\bm{r})$, where $\sigma=\{1,2,3\}$ corresponds to the particle generation label $\Sigma=\{\tau,\mu,e\}$.
        
   The rest frame linear and angular momentum vector potential for the fundamental fermions are
   	\begin{linenomath} 
	    \begin{subequations}\label{eq:VectorPotentialKS}
	        \begin{alignat}{1}
	            A^{[\rho:{\theta_m}]}_\mu & =-t\cdot\mathrm{diag}\big({\mathsf{K}}^{[\rho:{\theta_m}]}\big)^\gamma_\mu\partial_\gamma \phi ,\label{eq:VectorPotentialK} \\
	            {\tilde{A}}^{[\sigma:{\Theta_l}]}_\mu &= -t\cdot{{\mathsf{S}}^{[\sigma:{\Theta_l}]}}^\gamma \partial_\mu \partial_\gamma\varphi,\label{eq:VectorPotentialS}
	        \end{alignat}
		\end{subequations}
	\end{linenomath} 
    respectively, where ${\theta_m}$, $\mathsf{K}^{[\rho:{\theta_m}]}$ and $\phi$ are the color parameter, $\mathcal{K}$-chromospin vector and scalar potential of the particle's electric charge; while ${\Theta_l}$, $\mathsf{S}^{[\sigma:{\Theta_l}]}$ and $\varphi$ are the color parameter, $\mathcal{S}$-chromospin vector and scalar potential of the particle's spin. The momentum chromospin vectors and their corresponding scalar potential solutions are presented in Table \ref{tab:ColorConfigK} and in Table \ref{tab:ColorConfigS}. The angular momentum scalar potential has the unique value, $\varphi=1/R$, where $R=\sqrt{x^2+y^2+z^2}$. We use the ``[\textit{parent:child}]'' parametrization notation for the chromospin vectors to emphasize the dependency of color on the source charge and we suppress the color parameter when it is not required.	         
    
    \section{III. QCD color charge}
    The number of QCD color states for an elementary particle with a charge parameter, $\rho$, in three-dimensional space is ${}^3C_\rho$. For leptons, $\rho\in \{0,3\}$ and ${}^3C_{0,3}=1$, so we have only one color(less) state, whereas for quarks, $\rho\in \{1,2\}$ and ${}^3C_{1,2}=3$, so we have three distinct color states. These color states correspond to the independent spatial modes of the $\mathcal{K}$-chromospin vector, which is the union of linear momentum primitives (that reside on either side of an orientable 3-manifold: a slice of space-time), with algebraic form
	\begin{linenomath}      
	    \begin{subequations}\label{eq:doubleLayerK}
	    	\begin{alignat}{3}
	        {\mathsf{K}}^{[\rho:{\theta_m}]} ={\xi}^{[\rho:\theta_m]}_-  \uplus {\xi}^{[\rho:\theta_m]}_+,\\
	        {\mathsf{K}}^{[\bar{\rho}:{\theta_m}]} ={\xi}^{[\bar{\rho}:\theta_m]}_-  \uplus {\xi}^{[\bar{\rho}:\theta_m]}_+,
	        \end{alignat}
	    \end{subequations}
	\end{linenomath} 
    where $\xi^{[\bar{\rho}]}=-\xi^{[\rho]}$, $\xi^{[\rho/\bar{\rho}]}_- = \min(\xi^{[\rho/\bar{\rho}]},\xi^{[0]})$ and $\,\xi^{[\rho/\bar{\rho}]}_+ = \max(\xi^{[\rho/\bar{\rho}]},\xi^{[0]})$. 
    \begin{table}[h]
    	\caption{\label{tab:ColorConfigK} (Color online) The $\mathcal{K}$-chromospin vector and corresponding scalar potential solutions that satisfy Maxwell's equations. The zeroth element is ${\mathsf{\xi}}^{[0]}=0000\rangle$. We employ the complex representation $|\Re,\Im \rangle$ in all visualizations and also drop the time component, ``$0$'',  for clarity.}
    	\begin{ruledtabular}
    		\begin{tabular*}{0.5\textwidth}{cccc}
    			$\qquad {\theta_m}\,/\,\rho$ & ~~~$1$ & ~~~$2$&$3$ \\ \hline
    			${\mathsf{\xi}}^{[\rho:{\theta_m}]}\qquad$ &&&\\
    			$\qquad r$  & ~~~~$0\bar{1}00\rangle $ & ~~~$0011\rangle $ & $0\bar{1}\bar{1}\bar{1}\rangle$ \\
    			$\qquad g$  & ~~~~$00\bar{1}0\rangle $ & ~~~$0101\rangle $ & $0\bar{1}\bar{1}\bar{1}\rangle$ \\
    			$\qquad b$  & ~~~~$000\bar{1}\rangle $ & ~~~$0110\rangle $ & $0\bar{1}\bar{1}\bar{1}\rangle$ \\ \hline
    			$\phi^{[\rho:{\theta_m}]}\qquad$ &&& \\
    			$\qquad r$  & ~~~$|x|\delta^2(y,z)$~ & ~~~~$\ln \sqrt{y^2+z^2}~\delta(x)$ & $-1/R$ \\ 
    			$\qquad g$  & ~~~$|y|\delta^2(x,z)$~ & ~~~~$\ln\sqrt{x^2+z^2}~\delta(y)$  & $-1/R$ \\
    			$\qquad b$  & ~~~$|z|\delta^2(x,y)$~ & ~~~~$\ln \sqrt{x^2+y^2}~\delta(z)$ & $-1/R$ \\ \hline \\
    			\ifdefined \preprintFormat
    			\multicolumn{4}{c}{\includegraphics[scale=0.07]{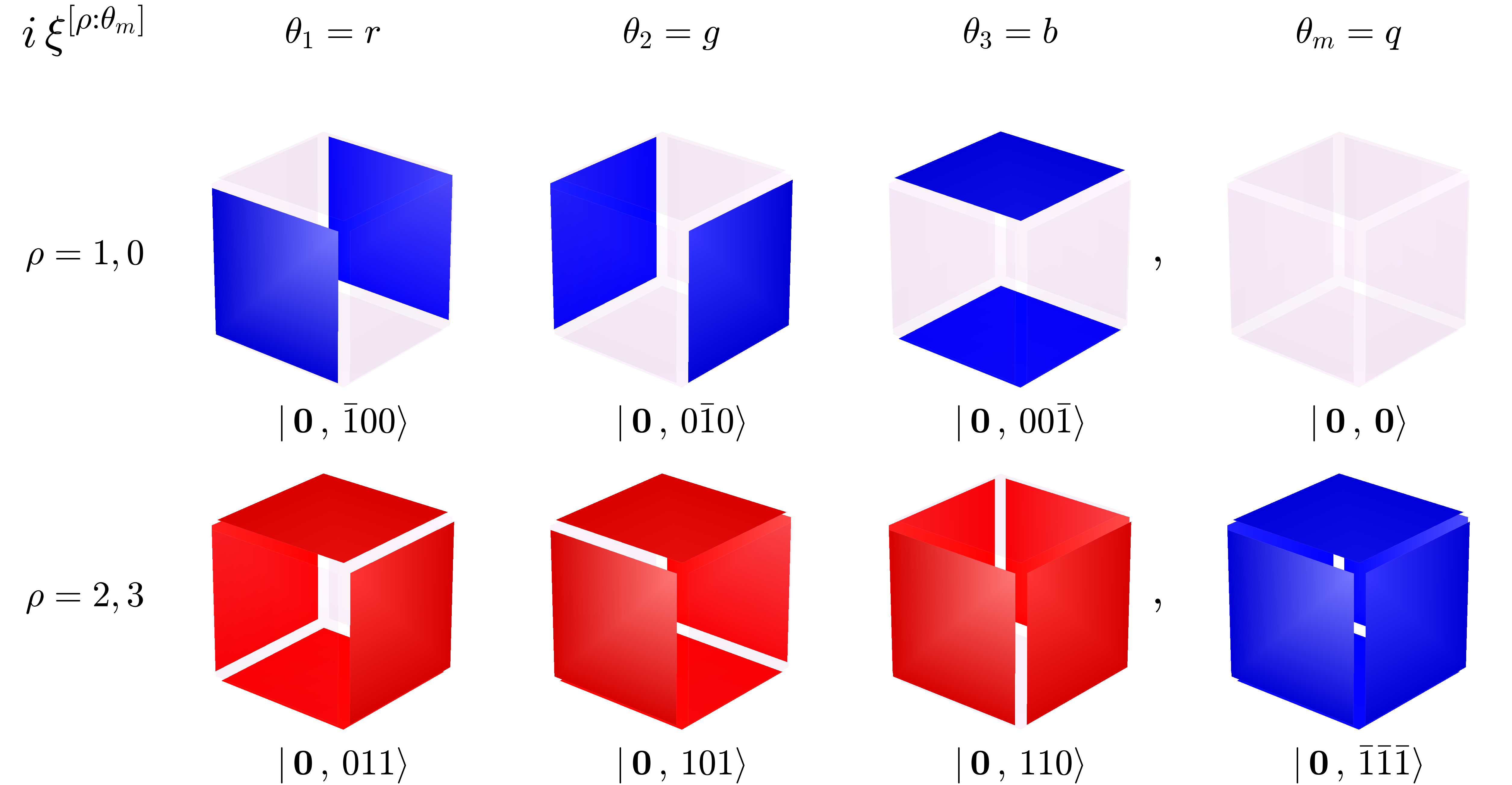}}
    			\else
    			\multicolumn{4}{c}{\includegraphics[scale=0.05]{Tab_II}} 
    			\fi                  
    		\end{tabular*}
    	\end{ruledtabular}             
    \end{table}
    The values of $\xi^{[\rho]}$ are presented in Table \ref{tab:ColorConfigK} and we note that its color assignment differs from the Han-Nambu scheme: where a one-to-one correspondence is made between the color states and the entries of the $\mathcal{K}$-chromospin vector. The $\mathcal{K}$-chromospin primitive, ${\xi}^{[\rho:{\theta_m}]}_\pm$, encodes the source parity and has the following occupancy ordering $|txyz\rangle$, where an entry value of zero implies that absence of the particle's electric field in that direction. The scalar potential solutions for the various boundary conditions and color states of the fermions are presented in Table \ref{tab:ColorConfigK}.

    \section{IV. Spin}
    The defining property of fermion spin in this framework is that any one of its bases element ($\mathcal{S}$-chromospin vector) span three-dimensional space. This can be inferred from the following quantum mechanical mapping: $\frac{1}{2}|\sqrt{\sigma^\complement+\sigma} ,\mathsf{S}_j\rangle \to |L,m_j\rangle$, where $\sigma^\complement = 3-\sigma$. The spin of a particle is encoded in its $\mathcal{S}$-chromospin vector---which is the union of the complementary $\mathcal{S}$-chromospin primitives, ${\zeta}^{[\sigma]}$ and ${\zeta}^{[\sigma^\complement]}$, that reside on either side of an orientable 3-manifold. The ``$l$-th'' color topological configuration of the $\mathcal{S}$-chromospin vector and its conjugate are:
	\begin{linenomath}      
	    \begin{subequations}\label{eq:SpinorCompletion}
	        \begin{alignat}{3}
	            &{\mathsf{S}}^{[\sigma:{\Theta_l}]} &&=  {\zeta}^{[\,\sigma\,\,:{\Theta_l}]} &&\uplus {\zeta}^{[\sigma^\complement:{\Theta_l}]},\label{eq:SpinorCompletionS} \\
	            &\tilde{\mathsf{S}}^{[\sigma:{\Theta_l}]} &&= {\zeta}^{[\sigma^\complement:{\Theta_l}]}  &&\uplus {\zeta}^{[\,\sigma\,\,:{\Theta_l}]},\label{eq:SpinorCompletionSbar}
	        \end{alignat}
	    \end{subequations}
	\end{linenomath}
    respectively, where $\Theta \supseteq \{\vartheta,\vartheta^\star\}$ and  ${\mathsf{S}}^{[\bar{\sigma}]}=-{\mathsf{S}}^{[\sigma]}$. 
    \begin{table}[h]
    	\caption{\label{tab:ColorConfigS} (Color online) The $\mathcal{S}$-chromospin vector and the corresponding chromospin tagged scalar potential from Eq. (\ref{eq:VectorPotentialS}). The ``$g$'' and ``$b$'' spin states are given by the left- and right-circular shift of the ``$r$'' spin state presented here and ${\zeta}^{[0]}=0000\rangle$.}
    	\begin{ruledtabular}
    		\begin{tabular}{c c c c}
    			$\qquad{\Theta_l}\,/\,\sigma$ & ~~~~~~~~$1$~~~~~~~~ &$2$&$3$ \\ \hline
    			${\zeta}^{[\sigma:{\Theta_l}]}\qquad$& & &\\
    			$\qquad r~\,$ & $0100\rangle$ & $00\bar{1}\bar{1}\rangle$ & $01\bar{1}\bar{1}\rangle$ \\ 
    			$\qquad r^\star$ & $0100\rangle$ & $0011\rangle$ & $0111\rangle$ \\ \hline
    			$\varphi_{,\gamma}^{[\sigma:{\Theta_l}]\gamma}\qquad$ &&&\\
    			$\qquad r~\,$ & $-x/R^3$ & $~~(y \oplus z)/R^3$  & $-(x \ominus y \ominus z)/R^3$ \\
    			$\qquad r^\star$ & $-x/R^3$ & $-(y \oplus z)/R^3$ &$-(x \oplus y \oplus z)/R^3$ \\ \hline \\
    			\ifdefined \preprintFormat
    			\multicolumn{4}{c}{\includegraphics[scale=0.07]{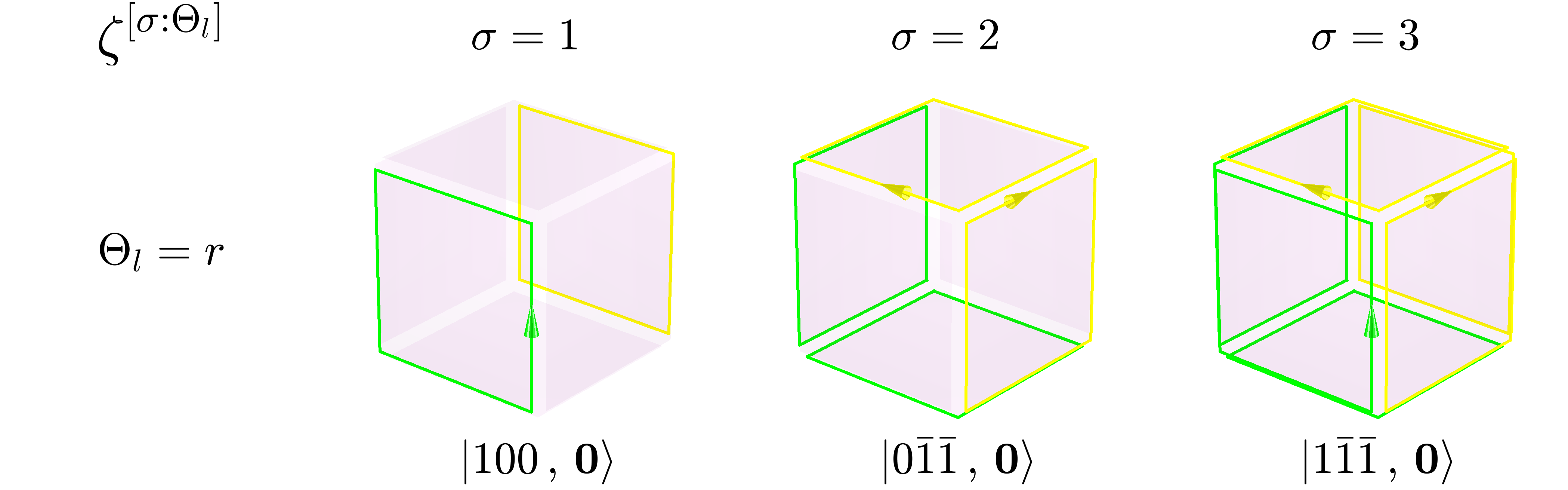}} 
    			\else
    			\multicolumn{4}{c}{\includegraphics[scale=0.05]{Tab_III}} 
    			\fi  
    		\end{tabular}
    	\end{ruledtabular}
    \end{table}
    Using the assigned values of the $\mathcal{S}$-chromospin vector in Table~\ref{tab:ColorConfigS}, ${\mathsf{S}}^{[\sigma:\vartheta_l]}$ corresponds to the spin-up state in the $x^l$-direction only while ${\mathsf{S}}^{[\sigma:\vartheta_l^\star]}$ corresponds to the spin-up state in all three spatial directions. To enumerate all possible spin values in a fixed color topology, we introduce the color triple, \{${\mathsf{S}}^{[\sigma:{\Theta_l}]}$, ${\mathsf{S}}^{[\sigma:{\Theta'_l}]}$, ${\mathsf{S}}^{[\sigma:{\Theta''_l}]}$\} where ${\mathsf{S}}^{[\sigma:{\Theta'_l}]} \Doteq {\mathsf{S}}^{[\sigma:{\Theta_{l+1}}]}$ and ${\mathsf{S}}^{[\sigma:{\Theta''_l}]} \Doteq {\mathsf{S}}^{[\sigma:{\Theta_{l+2}}]}$ represent the numerical equality of their three-dimensional spatial projections, whereas ${\mathsf{S}}^{[\sigma:{\Theta_l}]} \bumpeq {\mathsf{S}}^{[\sigma:{\Theta'_l}]} \bumpeq {\mathsf{S}}^{[\sigma:{\Theta''_l}]}$ represents their identical color topology. Note that the spin color index is confined to the ring of integers: $\mathbb{Z}/3\mathbb{Z}$. Finally, Eq. (\ref{eq:SpinorCompletion}) implies identical $\mathcal{S}$-chromospin configuration for particles with $\sigma=\{0,1\}$ and antiparticles with $\sigma=\{3,2\}$.
    
    \section{V. Fermion}
    We represent the fermions using a single momentum chromospin vector, $\psi$, which is made from the sum of the linear and angular momentum chromospin vectors. The chirality (and spin state) of a fermion derives from the real part of $\psi$ because,
    \begin{linenomath}
	    \begin{subequations}\label{eq:chromospinFermion}
	       	\begin{alignat}{3}
	       	&\mathsf{\psi}_{L/R} &&= {\mathsf{S}}^{[\bar{\sigma}/\sigma:\,\Theta]} && + i\,{\mathsf{K}}^{[\rho]},\label{eq:chromospinFermionLR}\\
	       	&\mathsf{\psi}^{(\downarrow/\uparrow)}_l &&= {\mathsf{S}}^{[\bar{\sigma}/\sigma:\,\Theta_l]} && + i\,{\mathsf{K}}^{[\rho]},\label{eq:chromospinFermionUD}
	       	\end{alignat}
	    \end{subequations}
	\end{linenomath}
    with chirality defined as the signed volume of the $\mathcal{S}$-chromospin vector, $\mathcal{C}=\epsilon^{ijk}\mathsf{S}_i\mathsf{S}_j\mathsf{S}_k$. It has a value of $-1(+1)$ for a $L(R)$ chiral fermion. With reference to Eq.~(\ref{eq:chromospinFermion}), $\psi_L + \psi_R$ encodes to electric charge information while $\psi_L - \psi_R$ encodes spin information.
        
    \subsection{A. Lepton}    
\begin{figure}[h]
    	\ifdefined \preprintFormat
    		\includegraphics[scale=0.115]{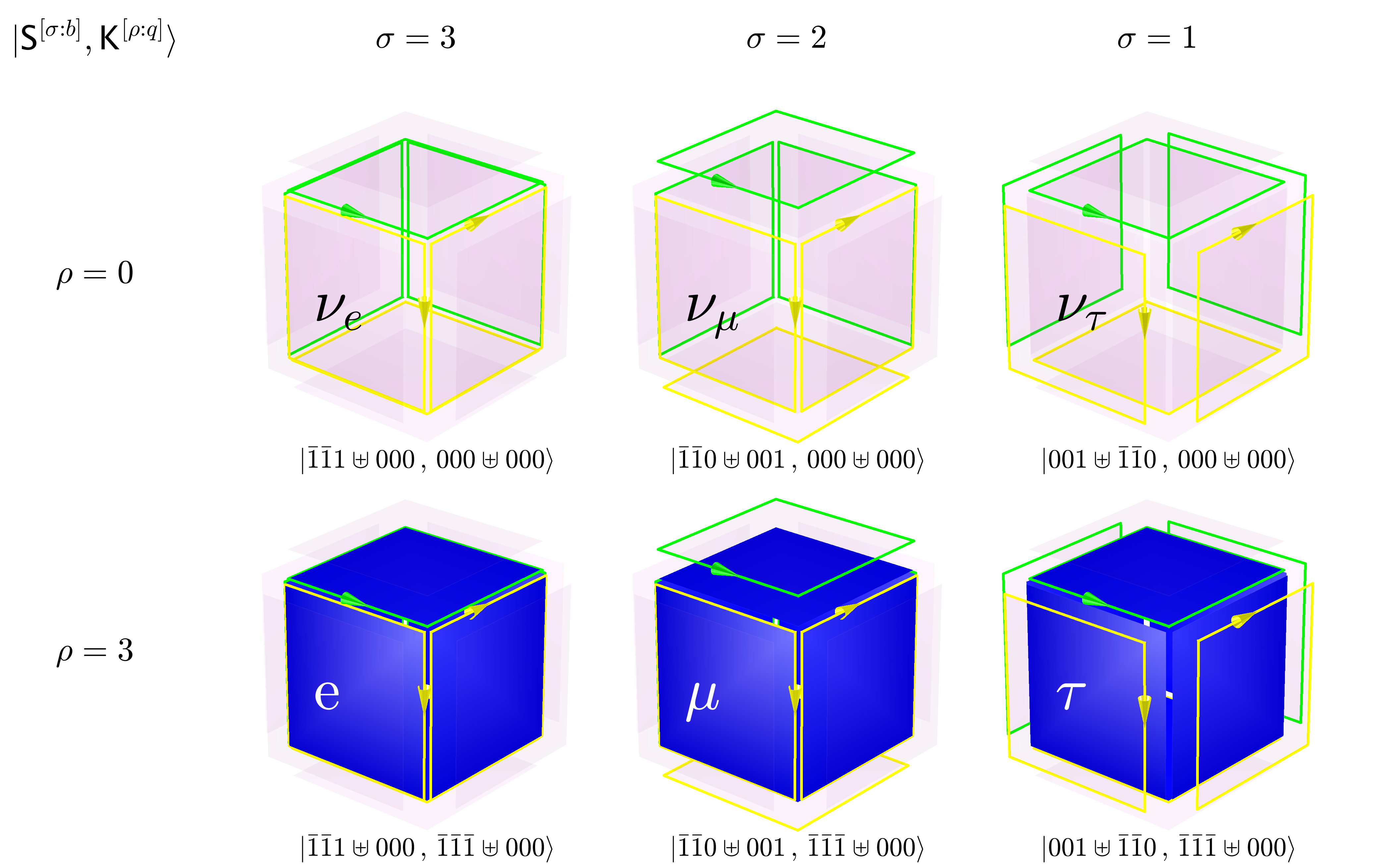}
    	\else
    		\includegraphics[scale=0.062]{Fig_1}
    	\fi        
    	\caption{ (Color online) Momentum chromospin representation of the leptons. We use the concentric shell model to illustrate current loops that reside on either side of an orientable 3-manifold. The spin configuration is chosen to minimize the overlap between the $\mathcal{K}$- and $\mathcal{S}$-chromospin vectors as we move from the electron to the tau generation. The chromospin overlap parameter, $\sigma$, governs a lepton's ability to radiate photons.}
        \label{fig:leptons}
    \end{figure}

	The momentum chromospin for a lepton and its anti-partner are:
    \begin{subequations}\label{eq:chromospinLeptonAntiLepton}
        \begin{alignat}{3}
            &\mathsf{\Lambda}^{[\sigma,\rho]}_n &&= {\mathsf{S}}^{[\sigma:\Theta_n]} &&+ i\,{\mathsf{K}}^{[\rho:q]},\label{eq:chromospinLepton}\\
            &\bar{\mathsf{\Lambda}}^{[\sigma,\rho]}_n &&= \tilde{\mathsf{S}}^{[\sigma:{\Theta}_n]} &&+ i\,{\mathsf{K}}^{[\bar{\rho}:q]},\label{eq:chromospinAntiLeptonVB}        
		\end{alignat}
    \end{subequations}
    respectively, where $\rho \in \{0,3\}$ for a \{neutral, charge\} lepton. Thus, the colorless state of the lepton is parametrized by either the presence or the absence of charge, $q$. Fig. \ref{fig:leptons} shows the three generations of leptons.
    With reference to Eq.~\ref{eq:chromospinLeptonAntiLepton}, the spin configurations of a particle and its anti-partner are  conjugate (in the topological sense) to ensure identical momentum chromospin overlap configuration for charge leptons. Furthermore, since the spin topology of $\nu_\mu (\bar{\nu}_\mu)$ and  $\bar{\nu}_\tau (\nu_\tau)$ are identical, their only distinguishing trait is that they correspond to opposite chirality states with reference to the weak interaction---as is the case in the SM. In this framework, the $W$ boson couples to the $L(R)$ charged(neutral) leptons and their $R(L)$ anti-patners. Neutrino oscillation \cite{neuosc1,neuosc2} shows that there is a non-zero probability for the spin excitation to tunnel between the two sides of the 3-manifold because of the degeneracy that occurs in the absence of charge excitation. Note that for the charged leptons, the excitation of all the charge sites precludes spin excitation tunneling and thus no mixing of the electron, muon and tau particles.
        
    \subsection{B. Quark}    
    The momentum chromospin for a quark and its anti-partner is parameterized by the middle range of the charge parameter, $\rho\in\{1,2\}$ and is given by:
    \begin{linenomath}
	    \begin{subequations}\label{eq:chromospinQuarkAntiQuark}
	        \begin{alignat}{3}
	            &\mathsf{\Omega}^{[\sigma,\rho]}_m &&= \tilde{\mathsf{S}}^{[\sigma:\Theta_m]} &&+ i\,{\mathsf{K}}^{[\rho:\theta_m]},\label{eq:chromospinQuark}\\
	            &\bar{\mathsf{\Omega}}^{[\sigma,\rho]}_m &&= {\mathsf{S}}^{[\sigma:{\Theta}_m]} &&+ i\,{\mathsf{K}}^{[\bar{\rho}:\theta_m]},\label{eq:chromospinAntiQuarkVB}       
			\end{alignat}
	    \end{subequations}
	\end{linenomath}
    where $m$ is the color index. For a quark in a fixed color state, the color triple \{${\mathsf{S}}^{[\sigma:{\Theta_l}]}$, ${\mathsf{S}}^{[\sigma:{\Theta'_l}]}$ ${\mathsf{S}}^{[\sigma:{\Theta''_l}]}$\} provides access to all the spin values, as would be the case during photon emission and absorption.   
    \begin{figure}[h]
        \ifdefined \preprintFormat
        \includegraphics[scale=0.115]{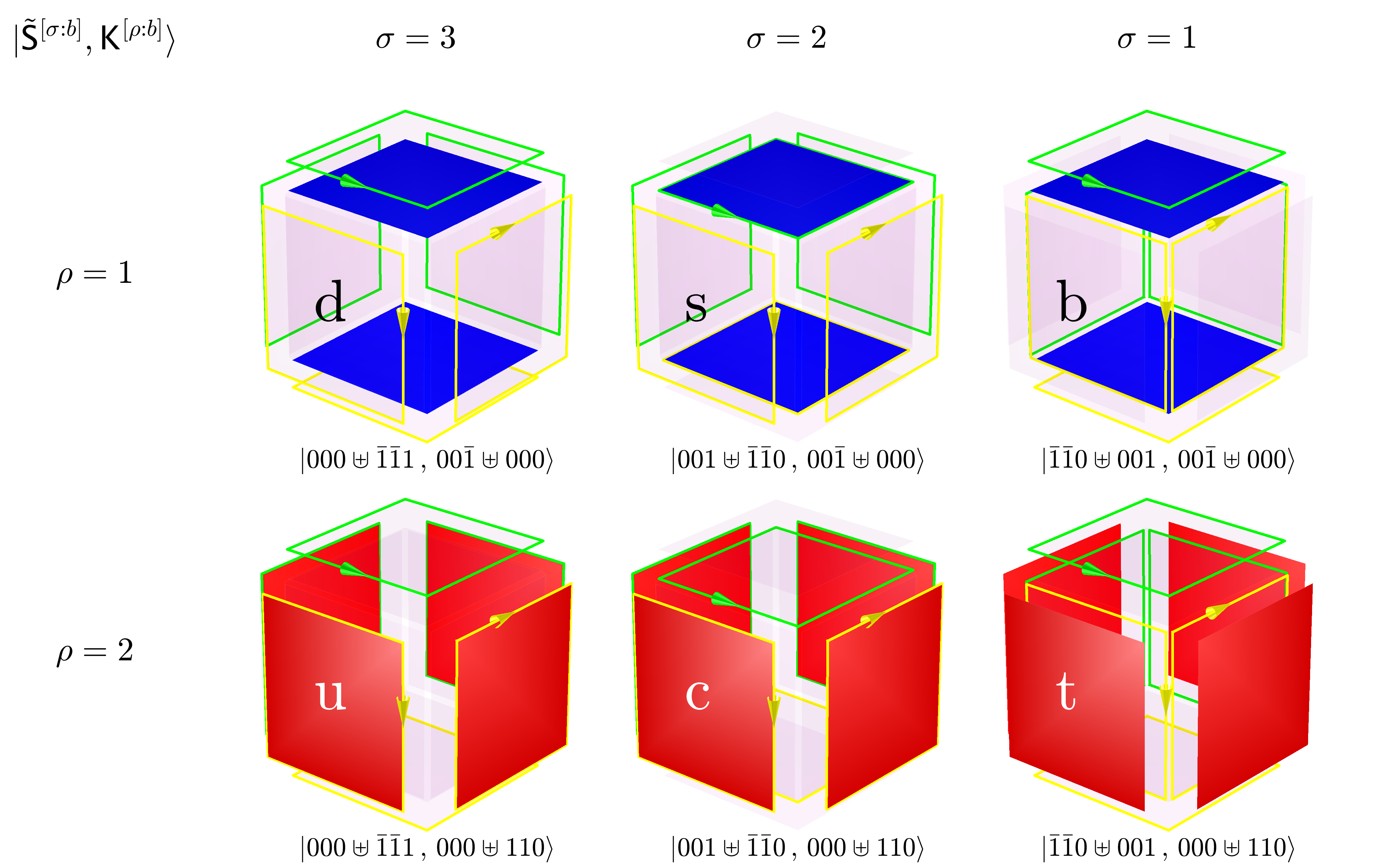}
        \else
        \includegraphics[scale=0.062]{Fig_2}
        \fi        
        \caption{  (Color online) Momentum chromospin representation of the quarks. The spin configuration of the quark is conjugate to the lepton in order to minimize the overlap between the $\mathcal{K}$- and $\mathcal{S}$-chromospin vectors as we move from the up quark to the top quark generation.}
        \label{fig:quarks}
    \end{figure}
    Fig.~\ref{fig:quarks} shows the graphical representation of the momentum chromospin vectors for the quarks \cite{quark}. The total electric field energy (scalar potential solutions in Table~\ref{tab:ColorConfigK} without the $\delta$ functions) is unbounded at infinity for the quarks . Therefore, quarks cannot exist in isolation but instead form bound states in order to eliminate the aforementioned energy pole at infinity, giving rise to quark confinement. Moreover, since only a fraction of the charge sites are excited in the quark sector, there is room for spin excitation tunneling that results in quark flavor mixing: which we expect to be fractionally smaller than neutrino mixing [see CKM and PMNS matrices].

    \section{VI. Fundamental bosons}
    The spin parameter ($\sigma$) of the [Higgs boson, Photon, Gluon, W/Z boson] has a value of $[0,1,2,3]$, which encodes the number of simultaneous spin flip (in the interacting fermion) that occurs during their emission or absorption \cite{RPP2018}. This process produces a $\sigma$\textit{-Vector} boson interaction with $\mathcal{S}$-chromospin value $\pm 2\zeta^{[\sigma]}$, where \textit{0-Vector} represents the scalar interaction of the Higgs boson \cite{HB1,HB2,HB3,HB4,HB5}, \textit{1-Vector} represents the face-like interaction of the photon \cite{PHOTON}, \textit{2-Vector} represents the edge-like interaction of the gluon \cite{GLUON} and \textit{3-Vector} represents the vertex-like interaction of the weak gauge bosons \cite{WB}.
    
    \subsection{A. Higgs boson}
    The Higgs momentum chromospin  is best described as a superposition of a left(right) particle and the corresponding right(left) antiparticle, i.e.
    \begin{linenomath}
	    \begin{equation}\label{eq:VacuumAndHiggs}
	    \mathsf{H} = \sum \Pi_{L/R} + \bar{\Pi}_{R/L}, \\
	    \end{equation}
    \end{linenomath}
    where $\Pi \supseteq \{\mathsf{\Lambda}, \mathsf{\Omega}, \gamma, \mathrm{g}, \mathsf{W}, \mathsf{Z} \}$. We use the one-dimensional and two-dimensional determinants to compute the chirality of the photon ($\gamma$) and gluon ($\mathrm{g}$) angular momentum chromospin respectively.
        
    \subsection{B. Photon and Gluon}
\begin{figure}[h]
		\ifdefined \preprintFormat
	  	\includegraphics[scale=0.11]{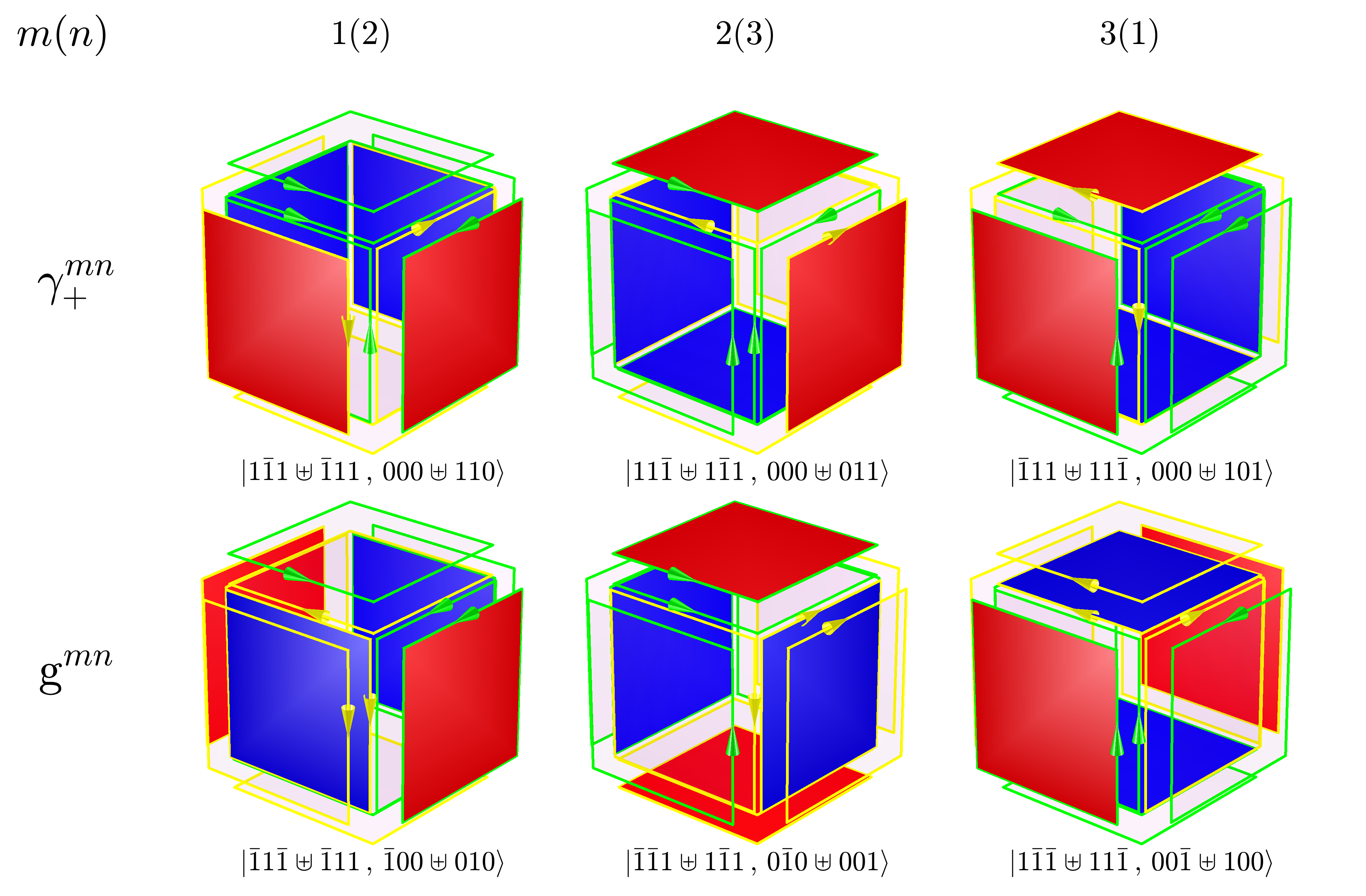}
	  	\else
	  	\includegraphics[scale=0.058]{Fig_3}
	  	\fi        
	  	\caption{ (Color online) Momentum chromospin representation of the photon and gluon in their transverse spin states.}
	  	\label{fig:Photon and Gluon}
	\end{figure}

   	The linear momentum chromospin vectors for the photon and gluon are
   	\begin{linenomath}
	   	\begin{subequations}\label{eq:PhotonGluonIsopinK}
	   		\begin{alignat}{3}
	   		&\mathsf{K}^{mn}_{\pm} &&= \mathsf{K}^{[1':{\theta_{m}}]} &&\pm \mathsf{K}^{[1':\theta_{n}]},\label{eq:PhotonIsopinK}\\
	   		&\mathsf{K}^{mn} &&= \mathsf{K}^{[1:{\theta_{m}}]} &&+ \mathsf{K}^{[\bar{1}:\theta_{n}]},\label{eq:GluonIsopinK}
	   		\end{alignat}             
	   	\end{subequations} 
   \end{linenomath}
   	respectively, where the ``\,$1'$\," notation implies that the photon $\mathcal{K}$-chromospin vector represents a dipole instead of the usual monopole. $\mathsf{K}^{12}_{\pm}$ represents the $\pm 45^\circ$ polarization state, $|\negthickspace\nearrow\rangle_{\odot z}$ or $|\negthickspace\nwarrow\rangle_{\odot z}$, of a photon traveling in the positive $z$-direction (out of page), whereas $\mathsf{K}^{12}$ describes the color state of a $r\bar{g}$ gluon. Let us build the angular momentum chromospin vectors, $2\zeta^{[1,2]}$, from the $\mathcal{S}$-chromospin of the electron generation as this provides an intuitive description of the fermion transformation that accompanies the exchange of a photon (or gluon). The transverse ($\triangle$) and longitudinal ($\Yup$) $\mathcal{S}$-chromospin vectors are:
   	\begin{linenomath}
	   	\begin{subequations}\label{eq:PhotonGluonIsopinS}
	   		\begin{alignat}{3}
			   	&\mathsf{S}^{mn}_{\pm} &&= \mathsf{S}^{[\bar{3}:\vartheta_m]} &&\pm \tilde{\mathsf{S}}^{[\bar{3}:\vartheta_n]},\label{eq:PhotonIsopinS} \\
			   	&\mathsf{S}^{l}_{\pm}  &&=  \mathsf{S}^{[3:{\vartheta_{l}]}}  &&\pm \tilde{\mathsf{S}}^{[3:\vartheta^\star_l]},\label{eq:GluonIsopinS}
	   		\end{alignat}
		\end{subequations}
	\end{linenomath}
   	respectively, where for example $\mathsf{S}^{12}_{+}$ ($\mathsf{S}^{3}_{+}$) is the transverse (longitudinal) spin transition, $\psi_{x}^{(\uparrow)} \to \psi_{y}^{(\downarrow)}$ \big($\psi_{xyz}^{(\uparrow)} \to \psi_{z}^{(\downarrow)}$\big), of any fermion that emits a photon that travels in the $z$-direction in a positive helicity state, $|\negthickspace\circlearrowleft\rangle_{\odot z}$, whereas $\mathsf{S}^{12}_{-}$ ($\mathsf{S}^{3}_{-}$) is the transverse (longitudinal) spin transition, $\psi_{x}^{(\uparrow)} \to \psi_{y}^{(\uparrow)}$ \big($\psi_{xyz}^{(\uparrow)} \to \psi_{z}^{(\uparrow)}$\big), of any quark that emits a $r\bar{g}$ gluon. Using Eq.~(\ref{eq:PhotonGluonIsopinK}) and Eq.~(\ref{eq:PhotonGluonIsopinS}), the momentum chromospin bases of the photon and gluon are
   	\begin{linenomath}
		\begin{subequations}\label{eq:Photon and Gluon}
			\begin{alignat}{3}
			&\gamma^{mn}_\pm &&=  \mathbb{S}^{mn}_{+} &&+ i\,\mathsf{K}^{mn}_{\pm},\\
			&\mathrm{g}^{mn} &&=  \mathbb{S}^{mn}_{-} &&+ i\,\mathsf{K}^{mn},
			\end{alignat}             
		\end{subequations}
	\end{linenomath}
	respectively, where $\mathbb{S}^{mn}_{\pm} \in \{\mathsf{S}^{mn}_{\mathrm{\pm}}, \,\epsilon^{lmn} \mathsf{S}^{l}_{\mathrm{\pm}}\}$. Therefore, $[\gamma^{12}, \gamma^{23}, \gamma^{31}]$ is the chromospin vector for a photon that is polarized in the $[xy, yz,zx]$-plane, while $[\mathrm{g}^{12}, \mathrm{g}^{23}, \mathrm{g}^{31}]$ is the chromospin vector for the $[r\bar{g}, g\bar{b},b\bar{r}]$ gluon. Fig.~\ref{fig:Photon and Gluon} shows a graphical representation of the momentum chromospin bases for the photon and gluon.  
	
	The photon chromospin solves Maxwell's equations with the following coordinate variable configurations: a $(t,x^k;x^i,x^j)$ partition for $A_\mu$ and a $(\tau;x,y,\check{z})$ partition for $\tilde{A}_\mu$. Thus, a $|\negthickspace\circlearrowleft, \nearrow \rangle_{\odot z}$ photon has the following momentum vector potentials:
	\begin{linenomath}
		\begin{subequations}\label{eq:VectorPotentialKPhoton}
			\begin{alignat}{1}
			{}^{\mathsf{\gamma}}A^{12}_\mu & = f(z-t)\cdot \big[0,1,1,0\big]^\nu\, \partial_\mu\partial_\nu \ln \sqrt{x^2+y^2}, \label{eq:VectorPotentialPhoton} \\
			{}^{\mathsf{\gamma}}\tilde{A}^{12}_{\check{\mu}} &= \tau\cdot \big[0,0,0,2\big]^{\check{\nu}}\, \partial_{\check{\mu}}\partial_{\check{\nu}} {\Big(1/\sqrt{x^2+y^2+\check{z}^2}\Big)},		
			\end{alignat}
		\end{subequations}
	\end{linenomath}
	where $f(z-t)$ is the usual envelop function for the classical EM field; $\tau=z+t$, $\check{z} = z-t$ and $\partial_{\check{\mu}} = -\big(\partial_\tau, \partial_x, \partial_y, \partial_{\check{z}}\big)$ are the light-cone coordinate variables that describe the angular momentum vector potential and its wave equation. With reference to Table~\ref{tab:ProtonQuarkOrbit}, gluon exchange provides the means to symmetrize the fractional charges of the quarks as illustrated by the proton's zero-point orbit (quark location and orientation). This process is periodic and occur at timescales small enough to suppress the isolated quark field solution. Also, the usual push-pull mode (photon exchange) of electromagnetism transmutes into the twist-swing mode (gluon exchange) of QCD at small distances---a manifestation of asymptotic freedom \cite{ASYM}.
	\begin{table}[h]
		\caption{\label{tab:ProtonQuarkOrbit} The QCD orbit of the valence quarks of the proton with an orbital parameter, $n$. The quarks occupy the vertices (indexed by  $j$) of a tetrahedron because this arrangement provides a stable three-quark configuration that enables \textit{2-Vector} gluon exchange. The exchange of a gluon between two quarks along any edge ($\triangledown$) of the tetrahedron is accompanied by the motion of the third quark along the corresponding dual edge ($\Ydown$) to the unoccupied vertex of the tetrahedron. These two processes (twist/swing) occur in tandem to create a stable orbit with a period of 12, i.e. $n \in \mathbb{Z}/12\mathbb{Z}$.}
		\begin{ruledtabular}
			\begin{tabular}{c|cccccccccccc}
				\ifdefined \preprintFormat
				 \multicolumn{13}{c}{\includegraphics[scale=.08]{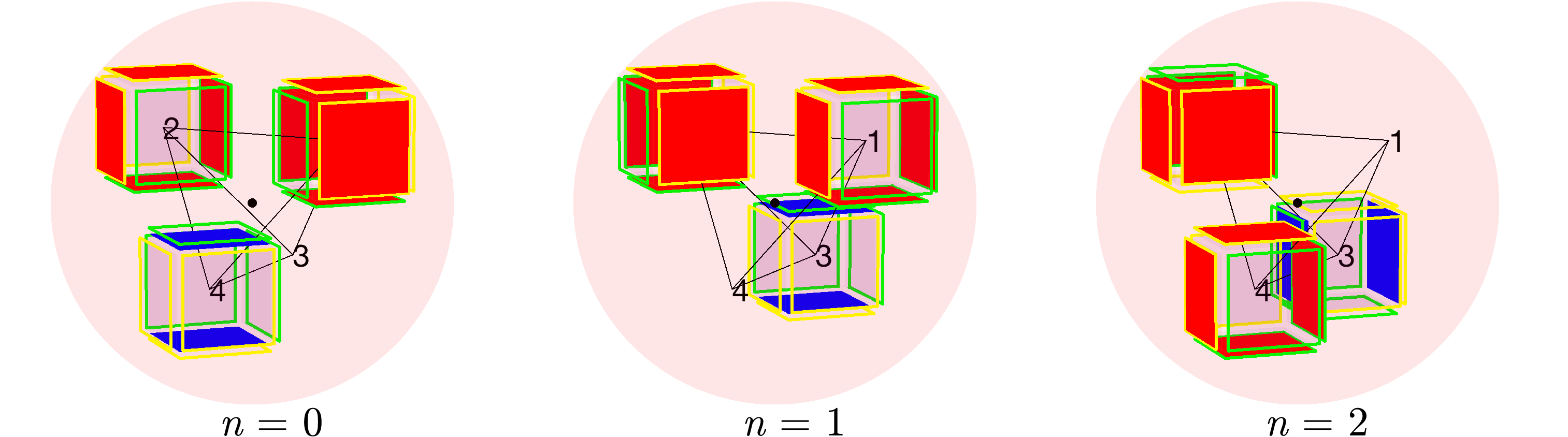}}\\ \hline
				\else
				\multicolumn{13}{c}{\includegraphics[scale=.06]{Tab_IV}}\\ \hline
				\fi        
				
				$j/n$& 0 & 1 & 2 & 3 & 4 & 5 & 6 & 7 & 8  & 9 & 10 & 11\\ \hline
				$1$&
				$\mathrm{u}_r$&$\mathrm{u}_g$&$-$&$\mathrm{u}_b$&$\mathrm{u}_r$&$\mathrm{u}_g$&$-$&$\mathrm{d}_b$&$\mathrm{d}_r$&$\mathrm{d}_g$&$-$&$\mathrm{u}_b$\\ \hline
				$2$ &$\mathrm{u}_g$&$\mathrm{u}_r$&$\mathrm{u}_b$&$-$&$\mathrm{d}_g$&$\mathrm{d}_r$&$\mathrm{d}_b$&$-$&$\mathrm{u}_g$&$\mathrm{u}_r$&$\mathrm{u}_b$&$-$\\ \hline
				$3$ &$-$&$\mathrm{d}_b$&$\mathrm{d}_r$&$\mathrm{d}_g$&$-$&$\mathrm{u}_b$&$\mathrm{u}_r$&$\mathrm{u}_g$&$-$&$\mathrm{u}_b$&$\mathrm{u}_r$&$\mathrm{u}_g$\\ \hline
				$4$ &$\mathrm{d}_b$&$-$&$\mathrm{u}_g$&$\mathrm{u}_r$&$\mathrm{u}_b$&$-$&$\mathrm{u}_g$&$\mathrm{u}_r$&$\mathrm{u}_b$&$-$&$\mathrm{d}_g$&$\mathrm{d}_r$			
			\end{tabular} 
		\end{ruledtabular}                    
	\end{table}

	\subsection{C. Weak gauge bosons}
	The gauge bosons for the weak interaction have the following momentum chromospin vectors:
	\begin{linenomath}
		\begin{subequations}\label{eq:chromospinWZ}
			\begin{alignat}{2}
			&\mathsf{Z}^{\sigma\rho}_n &&= \psi^{[\sigma,\rho]}_n +\bar{\psi}^{[\sigma,\rho]}_n,\label{eq:BosonZ} \\
			&\mathsf{W}^{\sigma\rho}_n &&= \psi^{[\sigma,\rho]}_n +\bar{\psi}^{[\sigma,\rho^\complement]}_n,\label{eq:BosonW}
			\end{alignat}
		\end{subequations}
	\end{linenomath}
	where $\psi^{[\sigma,\rho]}_n$ is the underlying fermion chromospin, $n$ identifies the spin polarization, $\rho$ enumerates the various charge particle channels and as usual $\rho^\complement=3-\rho$. We combine the relevant charge and generation channels into the physical $Z^0$ and $W^\pm$ boson states as follows:
	\begin{linenomath}
		\begin{subequations}\label{eq:chromospinWpmZZbar}
			\begin{alignat}{2}
			&\mathsf{Z}^0_n &&= \frac{1}{12}\sum_\sigma\sum_{\rho}\mathsf{Z}^{\sigma\rho}_n,\label{eq:BosonZZbar} \\
			&\mathsf{W}^{\pm}_n &&= ~\frac{1}{6}\sum_\sigma\sum_{\rho}^{\rho_\pm}\mathsf{W}^{\sigma\rho}_{n},\label{eq:BosonWpm}
			\end{alignat}
		\end{subequations}
	\end{linenomath}
	where $\rho_-=\{0,1\}$ and $\rho_+=\{2,3\}$. The \textit{3-Vector} interaction of the weak boson is responsible for its parity asymmetry because the natural orientation of a triple flip in the spin part of a fermion is along the $0111\rangle$ direction in the fermion's spin frame. Moreover, the charge asymmetry of the weak interaction require that fermions emit the $\rho_-(\rho_+)$ state in the  $ 0\bar{1}\bar{1}\bar{1}\rangle(0111\rangle)$ direction and that these states unfold (sliding operation) in the same way during disintegration.
	\begin{figure}[h]
		\ifdefined \preprintFormat
		\includegraphics[scale=0.108]{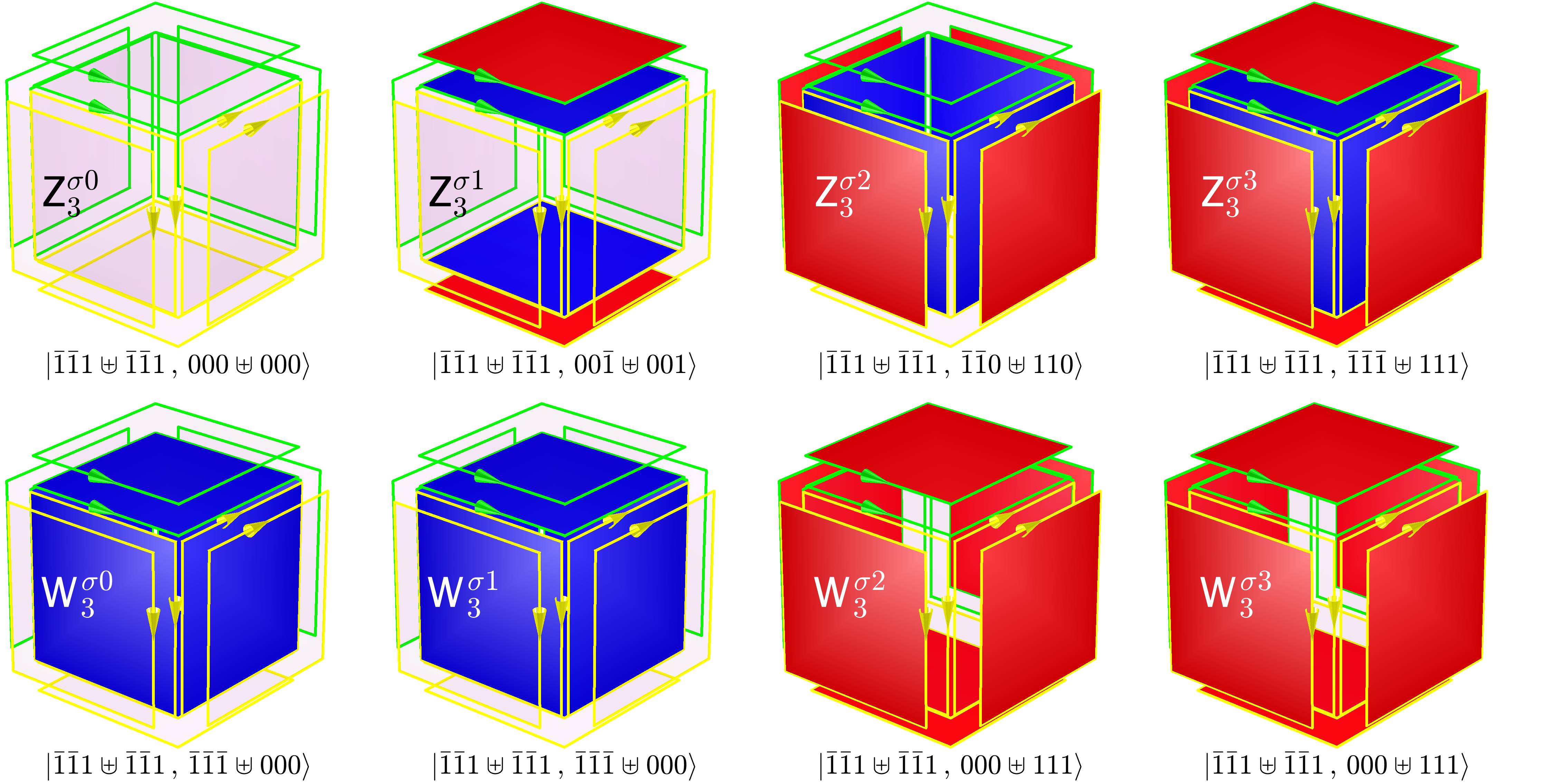}
		\else
		\includegraphics[scale=0.056]{Fig_4}
		\fi        
		\caption{ (Color online) Momentum chromospin of the {\it W} and {\it Z} bosons in the four different charge channels. The subscript, $n~(=3$ here), allows us to \textit{3-Vector} couple the weak bosons to all spin states of the fermion.}
		\label{fig:BosonWZ}
	\end{figure}
	
	The linear and angular momentum vector potential are ${}^{Z/W}\negthickspace A_\mu$ and ${}^{Z/W}\negthickspace\tilde{A}_\mu$ respectively. In this framework, the Yukawa type solution to the  Maxwell's equation imposes a finite lifetime on the $W/Z$ boson via a $e^{-\mu t}$ time dependence, where $t\geq0$ and $1/\mu$ is the mean lifetime. The scalar potential solution is $\phi,\varphi=e^{-\mu {(t+R)}}/R$, so that the linear and angular momentum vector potentials for the $W$  boson are:  ${}^{W}\negthickspace A_\mu = (0, \, \partial_j \phi)$ and ${}^{W}\negthickspace \tilde{A}_\mu = (0, \, 2\delta^k \partial_k\partial_j \varphi)$ respectively. The source dynamics suggests that the $W$ boson's creation is followed by a consolidation of its source moment of inertia prior to decay.
	    
    \section{Conclusion}
    In this paper, we develop the momentum chromospin model of the elementary particles which naturally resolves a number of outstanding particle physics problems including quark confinement, the duplication of fermions into three generations and a unified description of the fundamental fermions and bosons. We also illuminate the organizing principle that govern the properties of the four fundamental bosons and the role of charge (and spin) in fermion flavor mixing.

\end{document}